\documentclass[fleqn,10pt]{JLA_article} 

\usepackage[english]{babel} 
\usepackage{xcolor}

\usepackage{hyperref} 
\hypersetup{hidelinks,colorlinks,breaklinks=true,urlcolor=color2,citecolor=color1,linkcolor=color1,bookmarksopen=false,pdftitle={Title},pdfauthor={Author}}

\PaperTitle{FATE in MMLA: A Student-Centred Exploration of Fairness, Accountability, Transparency, and Ethics in Multimodal Learning Analytics} 

\Authors{Yueqiao Jin\textsuperscript{1}*, Vanessa Echeverria\textsuperscript{1}\textsuperscript{2}, Lixiang Yan\textsuperscript{1}, Linxuan Zhao\textsuperscript{1}, Riordan Alfredo\textsuperscript{1}, Yi-Shan Tsai\textsuperscript{1}, Dragan Gašević\textsuperscript{1}, Roberto Martinez-Maldonado\textsuperscript{1}} 
 \affiliation{\textsuperscript{1}\textit{Centre for Learning Analytics, Monash University, Clayton, Australia. Email: ariel.jin@monash.edu}} 
 \affiliation{\textsuperscript{2}\textit{Escuela Superior Politécnica del Litoral, 30.5 Via Perimetral, Guayaquil, Ecuador}}


\Submitted{xx/xx/xx}
\Accepted{xx/xx/xx}
\Published{xx/xx/xx}

\Volume{1}
\Number{1}
\Pages{1---10}
\Doi{xxx-xxx-xxx}
\Notesname{Notes for Practice} 

\note{Research in Fairness, Accountability, Transparency, and Ethics (FATE) can foster the design of Multimodal Learning Analytics (MMLA) systems that are equitable, responsible, and trustworthy, assuring that they benefit a broad spectrum of students without reinforcing biases or discrimination.}
\note{This paper presents the first evaluation of student perceptions on FATE in a large-scale MMLA deployment, providing empirical evidence from an in-depth qualitative study from a student-centred perspective to identify a set of specific FATE issues that need to be addressed in future MMLA studies and practices}
\note{The paper highlights the need for fair and transparent data representation, multi-level data access, clarity in data collection and analysis processes, and measurable informed consent, advocating for research and practices that are ethically robust, student-centred, and accountable in addressing FATE issues.}

\Abstract{Multimodal Learning Analytics (MMLA) integrates novel sensing technologies and artificial intelligence algorithms, providing opportunities to enhance student reflection during complex, collaborative learning experiences. Although recent advancements in MMLA have shown its capability to generate insights into diverse learning behaviours across various learning settings, little research has been conducted to evaluate these systems in authentic learning contexts, particularly regarding students' perceived fairness, accountability, transparency, and ethics (FATE). Understanding these perceptions is essential to using MMLA effectively without introducing ethical complications or negatively affecting how students learn. This study aimed to address this gap by assessing the FATE of MMLA in an authentic, collaborative learning context. We conducted semi-structured interviews with 14 undergraduate students who used MMLA visualisations for post-activity reflection. The findings highlighted the significance of accurate and comprehensive data representation to ensure visualisation fairness, the need for different levels of data access to foster accountability, the imperative of measuring and cultivating transparency with students, and the necessity of transforming informed consent from dichotomous to continuous and measurable scales. While students value the benefits of MMLA, they also emphasise the importance of ethical considerations, highlighting a pressing need for the LA and MMLA community to investigate and address FATE issues actively.}

\begin{document}

\flushbottom 

\maketitle 


\thispagestyle{fancy} 


\section{Introduction}

Multimodal Learning Analytics (MMLA) remains a relatively novel approach in learning analytics (LA), focusing on diverse data collection methods to examine complex learning processes \cite{blikstein2013mmla,worsley2021new}. Rather than solely using clickstreams and keystrokes, MMLA encompasses other modalities like gestures \cite{worsley2015leveraging}, body movements \cite{Zhao22}, eye movements \cite{schneider2018leveraging}, facial expressions \cite{sumer2021multimodal}, hand motions \cite{spikol2017estimation}, voice \cite{dmello2015multimodal}, and physiological responses \cite{azevedo2019analyzing}. Recognising the multifaceted nature of student interactions, MMLA offers insights for creating personalised, adaptive learning environments \cite{cukurova2020promise}. By leveraging diverse data, educators and researchers can design systems that address individual student needs while considering different factors that influence learning \cite{sharma2020multimodal}.

Despite the promise of MMLA, there are concerns about the technologies used and unforeseen practices that may arise \cite{prinsloo2023multimodal, cukurova2020promise, worsley2021new, ochoa_multimodal_2022}. Challenges include the intricacies of multi-channel modelling, such as mapping diverse data to educational constructs \cite{ochoa_multimodal_2022,martinez2020from}, and logistical issues like complex technical infrastructure impacting MMLA scalability \cite{yan2022scalability,ouhaichi2023rethinking}. Systematic reviews consistently point out socio-technical and ethical challenges when transitioning MMLA from research settings to classrooms \cite{prinsloo2023multimodal,yan2022scalability,Alwahaby2022,crescenzi2020multimodal}. Specifically, using advanced sensors in learning spaces may threaten student privacy \cite{prinsloo2023multimodal}. While more data modalities can refine modelling \cite{giannakos2019multimodal}, it does not guarantee fairer analytical results \cite{deho2022existing}. Additionally, research in authentic classrooms remains scarce, leaving a gap in understanding how to use MMLA effectively while addressing the emerging fairness, accountability, transparency, and ethics (FATE) concerns in Artificial Intelligence (AI) \cite{memarian2023fairness}. Despite being highlighted in the MMLA literature \cite{Alwahaby2022,cukurova2020promise,worsley2021new,Giannakos2022sensor-based}, authentic MMLA deployments have yet to thoroughly examine these FATE issues.

Researching FATE in LA and MMLA is vital \cite{memarian2023fairness, khalil2023fairness}. It can foster the design of LA systems that are equitable, responsible, and trustworthy, assuring that they benefit a broad spectrum of students without reinforcing existing biases or accentuating discrimination \cite{khalil2023fairness}. Similarly, studying FATE in MMLA is essential to identify the boundaries of MMLA in an authentic learning setting \cite{prinsloo2023multimodal}. Previous studies have emphasised that understanding students' perceptions of FATE issues can aid in informing the creation of analytics technologies and the strategies surrounding their ethical use \cite{Kasinidou2021Educating,memarian2023fairness} as well as addressing potential threats to acceptance and adoption \cite{hakami2020learning}. This paper makes inroads into this area by exploring FATE issues in the context of an authentic large-scale MMLA deployment. The contribution of this study is two-fold: 1) we presented the first evaluation of students' perceptions of FATE issues in an authentic MMLA deployment, and 2) we identified a set of specific FATE issues that need to be addressed in future MMLA studies. Addressing these issues is essential for ensuring MMLA provides a fair and accurate representation of learner data, respects learners' autonomy over their personal data, fosters learners' understanding of the data collection and analysis processes, and upholds the integrity of achieving measurable informed consent. 
\section{Background}

\subsection{Fairness, Accountability, Transparency and Ethics (FATE)}

Over the past decade, research and discussion around the FATE of data-intensive educational systems have become more prevalent in LA \cite{khalil2023fairness} and AI in education communities \cite{memarian2023fairness}. While LA and AI have brought forth numerous possibilities for optimising learning and improving educational outcomes, they also introduced or intensified several critical issues that could undermine the ethicality and beneficence of education, such as eliciting bias and discrimination and diminishing student privacy and agency \cite{ungerer2022ethical, holstein2019fairness, pardo2014ethical}. Before exploring prior FATE-related works, especially in MMLA, we first defined each FATE element by synthesising the findings of a recent systematic review of FATE in AI and higher education \cite{memarian2023fairness} and prior discussions of FATE in LA \cite{khalil2023fairness, holstein2019fairness, deho2022existing, slade2013learning, prinsloo2023multimodal}. 

\subsubsection{Fairness} The fairness of LA is crucial as it can affect decisions related to students' learning pathways, interventions, academic success, and overall well-being \cite{deho2022existing, holstein2019fairness}. This terminology is a complex concept involving both descriptive and technical dimensions \cite{memarian2023fairness}. Descriptively, fairness seeks to establish a landscape in education that rectifies unjust practices and mitigates biases, ensuring that algorithmic processes do not produce discriminatory outcomes \cite{shin2022seeing, islam2022incorporating}. From a technical viewpoint, fairness can be achieved through rigorous statistical and mathematical approaches during various stages of the machine learning pipeline, emphasising metrics such as statistical parity, equalised opportunity, and accuracy parity \cite{jiang2021towards, kim2022information}. Addressing fairness also involves recognising and tackling different types of biases and employing strategies ranging from pre-processing techniques to post-process approaches \cite{barbierato2022methodology}. In the context of LA, fairness should extend beyond ensuring the analysis and algorithmic processes result in unbiased and nondiscriminatory analytics \cite{holstein2019fairness}, but also assuring the equitable and impartial reporting and visualisation of these analytics in dashboards or other user-facing interfaces \cite{verbert2020learning}.

\subsubsection{Accountability} This is a multifaceted concept that encloses the responsibility and answerability of individuals, institutions, or systems for actions, decisions, and outcomes they produce \cite{memarian2023fairness}. Specifically, some researchers associated accountability with explicability and considered a system or process that can be explained as accountable \cite{bezuidenhout2021does}. Others see accountability as a measure to hold providers of automated systems answerable for their algorithmic decisions, including both system developers and data suppliers \cite{pagallo2017automation}. This unclarity of who should be held accountable is further complicated by the debate between human accountability and human-AI shared accountability when considering the implications of AI systems in education \cite{memarian2023fairness}. In this sense, the accountability of LA is beyond a particular individual or party but distributed among multiple relevant stakeholders \cite{prinsloo2017elephant}, including but not limited to institutions, teachers, students, researchers, and technology providers. For example, the technology infrastructure department of institutions may be held accountable for data security, while developers of the algorithms and dashboards might be responsible for ensuring algorithmic and reporting fairness \cite{pardo2014ethical}. In MMLA, accountability is also associated with data access as these analytics could be used by different stakeholders for multiple purposes (e.g., feedback, reflection, and potentially assessment) \cite{kasepalu2021teachers, mangaroska2021challenges}. Therefore, understanding access to learner data is essential for holding different parties accountable.

\subsubsection{Transparency} Transparency in data-driven technologies is closely linked to accountability; for these technologies to be accountable, they must be transparent \cite{tsai2020empowering}. From an algorithmic point of view, transparency means algorithms are understandable, either in technical terms or simpler language \cite{ungerer2022ethical}. Whereas, from a process-oriented aspect, it extends to the clarity around data usage, consent mechanisms, and recourse options, aligning with standards like Europe's General Data Protection Regulation (GDPR) \cite{regulation2018general}. \citeA{chaudhry2022transparency} introduced a transparency index for educational AI systems, classifying it into three tiers: transparent to algorithmic experts (e.g., AI researchers and practitioners), transparent to domain experts (e.g., educational technology experts and enthusiasts), and transparent to educational stakeholders (e.g., educators and students). Based on these prior works, transparency in LA could be defined as the articulation and comprehension of data-driven insights, methods, and the implications of analytics-driven interventions, ensuring that stakeholders, particularly students, can understand, trust, and engage with the analytics \cite{tsai2020empowering, ungerer2022ethical}.

\subsubsection{Ethics} Ethics in LA is a broad concept that focuses on the principles and values guiding the collection, analysis, and use of data, which often overlaid with some aspects of fairness, accountability, and transparency \cite{ungerer2022ethical}. Specifically, ethical research in LA often focuses on investigating issues related to privacy, informed consent, data security, and the broader impact of analytics on society \cite{tzimas2021ethical, rubel2016student}. Of which, informed consent is critical and a prerequisite for conducting any LA research as such mechanisms are essential for ensuring student privacy and autonomy \cite{slade2013learning}. Achieving informed consent in practice is often complex \cite{prinsloo2023multimodal}. For example, when providing students with explanatory statements and asking for their consent, it often implies their awareness of the collection, usage, and storage of their data \cite{tsai2020privacy}. However, given the length of these statements, it is probable that students did not read these statements and consented without being fully aware of the situation \cite{prinsloo2016student}. Additionally, in MMLA studies, the use of advanced sensing technologies adds another layer of complexity, such as students' level of comprehension regarding how their data will be collected, processed, and analysed by these sensors (e.g., positioning tracking, computer vision, and wearable biometrics sensors) \cite{prinsloo2023multimodal,beardsley2020enhancing}. This issue becomes particularly vital in low-risk MMLA research that adopts an opt-out consenting approach, whereby participants are given explanatory statements and need to actively opt-out to withdraw their participation and data \cite{junghans2005recruiting}. Understanding the intricacies of opt-out consenting is essential for advancing the ethicality of LA \cite{sun2019s}.

\subsection{FATE in Multimodal Learning Analytics}

MMLA has advanced rapidly over the past decades with the overarching goal of extending LA research not only beyond computer-mediated contexts but also to hybrid and physical learning settings \cite{ochoa_multimodal_2022, cukurova2020promise}. Specifically, this subfield of LA strives to derive data-driven insights concerning learners' metacognitive and emotional states, alongside their learning behaviours, by utilising fine-grained physical and physiological signals. \cite{blikstein2013mmla, ochoa_multimodal_2022}. This endeavour has become increasingly realised with the advancement in sensing technologies, such as computer vision algorithms for gesture and posture detection, wearable positioning tracking, and physiological sensors \cite{crescenzi2020multimodal, Giannakos2022sensor-based, sharma2020multimodal}. While the integration of advanced sensing capabilities has empowered MMLA with the ability to capture a range of different modalities (e.g., video, audio, heart rate, positioning, and gaze \cite{cukurova2020promise, worsley2021new}), many concerns have also been raised regarding the FATE of MMLA \cite{Alwahaby2022, yan2022scalability}. For example, in a systematic literature review of MMLA research, \citeA{Alwahaby2022} identified several ethical issues around the design and use of MMLA systems, including privacy, informed consent, data management, and ethical clearance. While these issues have been briefly mentioned, either in the introduction, background, or discussion sections, they have yet to be formally investigated as the main research objective of an MMLA study. Likewise, in another review, \citeA{yan2022scalability} further illustrated the lack of understanding of the fairness and the potential risks associated with existing MMLA solutions. Both these reviews advocated for the immediate need for FATE research in MMLA, especially from a human-centred instead of an algorithm-focused perspective, which has been the primary focus of prior FATE research regarding data-intensive educational systems \cite{khalil2023fairness, memarian2023fairness}. 

Of the few studies that have investigated FATE-related issues in MMLA, most were conducted in controlled laboratory settings and have yet to evaluate MMLA solutions in authentic learning settings. For example, \citeA{kasepalu2021teachers} investigated teachers' trust in an MMLA dashboard for monitoring students' collaboration and identified a relationship between trust and the level of transparency in data processing. In their study, teachers did not actually use the dashboard in practice. Instead, they were shown an 8-minute video of a 30-minute recording and evaluated the dashboard accordingly. Likewise, \citeA{mangaroska2021challenges} evaluated students' perspectives on the accessibility of their multimodal data collected during a lab study. The results illustrated a mixed perspective on data sharing, where some students were comfortable with sharing anonymous and aggregated data with educators, while others demonstrated concerns over the unequal power relations that multimodal data may introduce, potential usage in data profiling, and pervasive surveillance, especially with physiological data. However, the study \cite{mangaroska2021challenges} focused merely on the data collection process without illustrating and evaluating MMLA with students, which could affect the outcomes as the utility of these data remains unclear for students. To the best of our knowledge, none of the existing MMLA studies have evaluated students' perceptions of the FATE of an MMLA solution after the practical implementation in authentic learning settings, where the MMLA solution has actually been used by students during learning activities. Understanding these perceptions is crucial. It can provide valuable insights for future MMLA research, aiming to create solutions that are fair, accountable, transparent, ethical, and, most importantly, beneficial for enhancing the learning experience \cite{gavsevic2015let}. 

This study addresses this gap by examining students' perceptions of FATE issues regarding the use of MMLA visualisations in an authentic collaborative learning setting. Specifically, the following research questions were investigated: In an authentic MMLA deployment, \textbf{RQ1}-- how is \textit{fairness} perceived by students, especially regarding the representation of their data? \textbf{RQ2}-- how is \textit{accountability} perceived by students, especially regarding the access of their data? \textbf{RQ3}-- how is \textit{transparency} perceived by students, especially regarding the data collection and analysis processes? \textbf{RQ4}-- how is \textit{ethics} perceived by students, especially regarding the opt-out informed consent?

\section{Methods}
\subsection{The Authentic Learning Context}
\label{sec:learning-context}
The study was conducted in the context of high-fidelity healthcare simulations that enable students to practise prioritisation and teamwork skills. In these, students play the role of nurses and address a clinical problem while focusing on patient care management. Third-year students in the Bachelor of Nursing at [hidden] University engaged in these simulations, each spanning between 20 and 30 minutes. Students are grouped into teams of four, taking the roles of (2) primary and (2) secondary nurses. The scenario involves three phases: \textbf{Phase 1} -- Two primary nurses receive handover information for four patients, with one patient deteriorating. \textbf{Phase 2} -- Two secondary nurses are called to assist with patient care. \textbf{Phase 3} -- Following a Medical Emergency Team (MET) call by a nurse, a teacher, acting as a doctor, offers support. Immediately after the scenario, all students engaged in a teacher-guided debrief session.

\subsection{The MMLA System}
\label{sec:mmla-system}
A multimodal LA system [\href{http://}{\textit{code available upon acceptance of the paper}}] was used during the debrief of the simulations. 
This comprises: 1) the data collection component, which captures, synchronises and stores data from different sensors and devices (i.e., audio, positioning, video) (see Section \ref{sec:mdc}); and
2) the visualisation component, which is responsible for generating analytics and visualisations (see Section \ref{sec:visualisations}).

\subsubsection{Multimodal Data Collection}
\label{sec:mdc}
Each student received a belly bag with a positioning sensor to monitor their \textbf{x-y coordinates} and \textbf{body orientation} in the learning space. They also wore a wireless headset microphone for \textbf{audio} recording. Students were assigned colours according to their roles: red and blue for primary nurses (PN) 1 and 2, and green and yellow for secondary nurses (SN) 1 and 2. These helped in creating anonymous data streams. Additionally, a 180-degree camera was used to \textbf{video}-record the simulations and an \textbf{observation tool} enables teachers to tag and annotate relevant phases and events.

\subsubsection{Visualisations}
\label{sec:visualisations}
\begin{figure}[h]
\centering
\includegraphics[width=.99\textwidth]{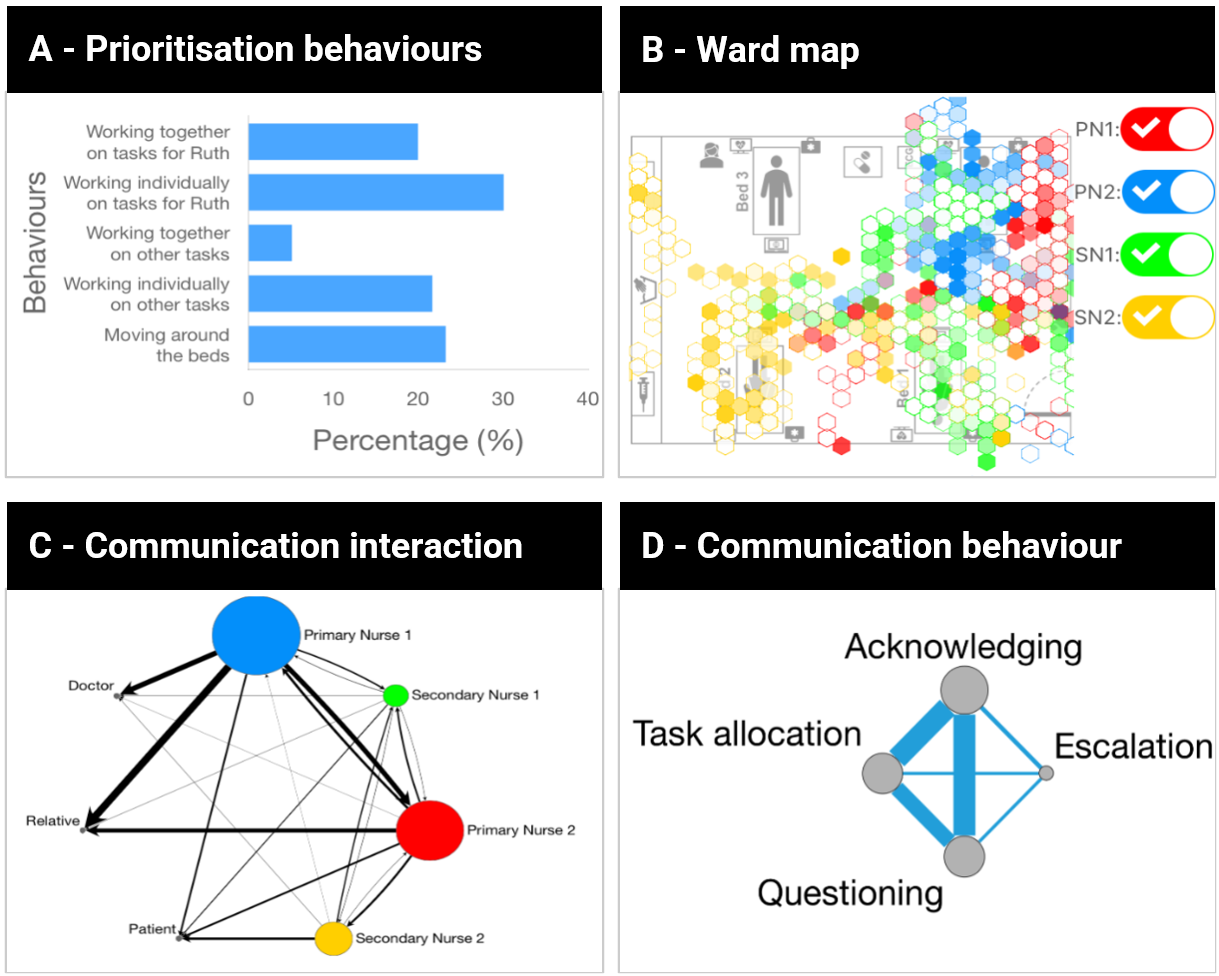}
\caption{A set of four visualisations used by teachers to guide reflection on collaboration dynamics in the classroom based on a combination of positioning (A, B, C and D) and audio (B, C and D) data. }
\label{fig:figure1}
\end{figure}

Teachers used a set of four visualisations to prompt reflections right after the simulations, capturing task prioritisation, teamwork, and communication behaviours (see Fig. \ref{fig:figure1}). Over nine months, the MMLA visualisations were designed through a co-creation with nursing teachers and a research team that consisted of software developers, interaction design experts, and LA researchers, guided by LATUX workflow \cite{martinez-maldonadoLATUXIterativeWorkflow2016}.
In a focus group discussion, we began by \textbf{\textit{exploring opportunities and challenges}} on how LA can be integrated into their current teaching practice to support post-simulation reflection. In the second stage, the teaching team generated \textbf{\textit{low-fidelity paper-based prototypes}}. In the third stage, the research team developed and tested \textbf{\textit{higher fidelity prototypes}} collaboratively and iteratively with teachers through three \textbf{\textit{pilot studies}} over six months. These visualisations were then integrated with the existing LA system in the physical classroom and used in 64 simulation sessions. The details about the visualisations will be open-sourced in our repository [\href{http://}{ hidden for review}].

Furthermore, we evaluated the visualisations with students using the Evaluation Framework for Learning Analytics (EFLA) for individuals’ understanding of data, awareness, reflection, and the impact on the learner through an 8-item questionnaire \cite{scheffel2017proof}. Complete results were detailed in our previous study \cite{yan2024evidence}, revealing that students perceived visualisations as effective in providing clarity on the data collected, stimulating reflection on, and adaptation in their learning behaviours. Additionally, they highlighted the importance of data accuracy, transparency, and privacy protection to maintain user trust. Therefore, in this paper, we emphasise exploring student reflections on FATE through these visualisations of their data.

\textbf{A--Prioritisation chart:} 
Based on students' \textit{\textbf{position data}}, and drawing from proxemics literature \cite{hall1971proxemics} and teacher insights \cite{yan2022indoor}, we identified five prioritisation behaviours. We checked if \textbf{\textit{students were collaborating}} by seeing if two or more were close (within 10 metres) for over 10 seconds. Otherwise, they were deemed \textit{\textbf{working individually}}. We also discerned if students were involved in a \textit{\textbf{primary}} or \textbf{\textit{secondary task}} by dividing the learning space into respective areas. The primary task pertained to the patient named "\textit{\textbf{Ruth}}", who needed prioritisation, while the secondary involved the other three patients. Finally, we tracked if students \textit{\textbf{moved around the beds}} when away from the main beds or other students. Using x-y coordinates, we tagged each student's behaviour every second into one of the five behaviours. We then aggregated the time the team dedicated to each behaviour, visualised in a bar chart (see Fig. \ref{fig:figure1}--A).

\textbf{B--Ward map:} 
We integrated \textit{x-y} coordinates with students' speech presence/absence data. Using a voice activity detector (VAD) \footnote{\url{https://github.com/wiseman/py-webrtcvad}}, we labelled speech utterances from individual students' audio recordings, producing timestamped instances indicating speech presence or absence. These instances were merged with \textit{x-y} coordinates to determine each student's location per second. The final matrix for each student included timestamps, audio presence (0/1), and x-y coordinates.
We depicted these spatial and audio data on a hexbin map\footnote{\url{https://d3-graph-gallery.com/hexbinmap.html}} to display the spatial-audio relationship. The \textit{x-y} coordinates were adjusted to fit the physical learning space layout. Data points were grouped in hexagons, each representing an area in the ward map. Students were colour-coded (Section \ref{sec:mdc}), with colour intensity reflecting speech activity. As shown in Fig. \ref{fig:figure1}--B, a \textbf{\textit{fully filled hexagon}} signified active speech, while \textbf{\textit{no fill}} denoted silence.

\textbf{C--Communication interaction diagram:} 
We combined students' positioning and audio data to visualise communication interactions. We identified f-formations – when two or more students were physically in close proximity – using individual x-y coordinates and body orientation. Using the VAD, we extracted speech segments and their lengths. We inferred conversations between students or other participants (e.g., doctor, patient, relative) if speech happened within an f-formation \cite{Zhao22}. For example, if PN1 (red) spoke within proximity of PN2 (blue), we assumed PN1 was conversing with PN2.
We computed each student's speaking time and visualised interactions through a node-link graph. As illustrated in Fig. \ref{fig:figure1}--C, nodes represent participants engaged in the simulation, and arrows illustrate the directionality of speaking. The node size represents the student's total speaking time, while the arrow width represents the interaction duration between two students. Nodes were colour-coded to preserve de-identification. 

\textbf{D--Communication behaviours diagram:} 
We adapted a coding scheme on healthcare teamwork and communication behaviours \cite{zhao2023mets} to analyse students' dialogue content. This was achieved by combining audio and positioning data. We used the VAD to identify when students spoke in the audio, providing us with corresponding timestamps. These timestamps were used to extract the voiced audio segments, which were subsequently transcribed using the OpenAI Whisper large model \footnote{\url{https://github.com/openai/whisper/tree/main}}. Moreover, by combining the timestamps and positioning data, we identified moments when two or more students conversed while in close proximity, forming dialogue segments. We then coded each utterance (i.e., a turn of talk) within a dialogue segment using the adapted coding scheme. Following a similar approach used by \citeA{zhao2023mets}, each utterance can be assigned with multiple codes. For visual representation, we developed a simplified epistemic network graph as shown in Fig. \ref{fig:figure1}--D. This network was implemented referring to the data processing algorithm of epistemic network analysis \cite{shaffer2016tutorial}. Each node symbolizes a code, and the lines between them indicate co-occurrences of codes within students' dialogue segments (i.e., similar to in Epistemic Network Analysis \cite{shaffer2016tutorial}). The thickness of the lines represents the extent of co-occurrences observed throughout the simulation.

\subsection{Study Design}

This study, approved by the [Institution Name] Human Research Ethics Committee, spanned four weeks in 2023's second semester as part of the [hidden] Unit. We conducted a follow-up semi-structured interview to delve into FATE within MMLA visualisations used in debriefs, structured as: \textit{i) Initial Explanation:} Despite prior exposure, we first clarified the data in each visualisation. \textit{ii) Understanding of the Information:} Students interpreted their data visualisations with opportunities for clarification. \textit{iii) FATE Questions:} Introducing a potential assessment tool scenario, we prompted discussion on LA's educational future. Queries covered Fairness (e.g., \textit{"Do you think all students in your team are fairly represented in the visualisations?, How?"}), Accountability (e.g., \textit{"Any concerns about data misuse in these visualisations?"}), Transparency (e.g., \textit{"Do you understand the data collection and transforming process?"}), and Ethics (e.g., \textit{"Would you prefer the 'opt-in' or 'opt-out' consent method? Why?"}). These questions drew from FATE research in HCI, LA, and AIED \cite{mangaroska2021challenges, prinsloo2023multimodal, hakami2020learning, khalil2023fairness, memarian2023fairness}. The complete list is available at the provided \href{https://anonymous.4open.science/r/FATE_in_MMLA-DC16/}{\textit{link}}.

\subsection{Participant Recruitment}
At the request of the teaching team, an opt-out consent method was adopted as the MMLA system became an integral part of all regular teaching sessions. An explanatory statement was provided to students, including the essential information about the study. Position and audio data were collected from all students, deleting those who opted out immediately after their sessions. Data from 60 teams (each comprised of four students) was captured during the simulations. Typically, each teaching session allows two teams to participate in the simulation. Teachers then utilised the visualisations for the debrief (Section \ref{sec:visualisations}), so students viewed these visualisations during their team discussions.
After class, students were invited to participate in a semi-structured interview to explore the research questions using their data. An opt-in consent approach was chosen for the interviews, and participants received a \$40 voucher as appreciation for their time. Conducted via Zoom, these interviews were video recorded and took place a week after the simulation experience, with a single researcher facilitating each session. Following the principle of saturation and guided by \citeA{guest2006many}'s recommended sample size, we recruited 15 students for the interviews. After conducting a pilot session and refining our questions based on feedback, the analysis focused on responses from 14 students (12 females, avg. age: 22 years, std. dev: 1.6) who participated in the subsequent interviews.

\subsection{Analysis}
All interviews were transcribed for analysis. We conducted a deductive-inductive thematic analysis approach \cite{fereday2006demonstrating}. First, using a deductive method, we sorted instances based on FATE concepts. Then, we used an inductive method to identify emerging themes within each concept. One researcher analysed the first three interviews following this approach. A second researcher then reviewed the findings, and both discussed to agreed on the identified themes \cite{mcdonald2019reliability}. After this, they analysed the remaining interviews in an iterative manner to reach saturation of themes and full agreement \cite{ando2014achieving}.

\section{Results}

\subsection{Fairness (RQ1)}

Students' perspectives on the fairness of the MMLA visualisations in reflecting their individual and team performance were mixed: six participants (P1, P3, P4, P9, P10, and P12) agreed; four (P2, P7, P8, and P11) disagreed; and four (P5, P6, P13, and P14) found the visualisations to be "somewhat fair". Their responses comprised four themes, covering positive and negative student perspectives of fair representation and potential improvements for assessment.

\subsubsection{Accuracy of Reflection}
\label{sec-fairness-accuracy}

The participants (n=7) felt that the MMLA visualisations accurately reflected simulation events. Those who saw them as a \textit{fair} representation noted a clear alignment between the visualisations and their experience. P1 felt it was an \textit{"accurate representation of what happened,"} while P3 confirmed the visualisation's representation of roles, observing that \textit{"Secondary Nurse 2 had the smallest role to play... that was also true within the simulation."} P4 mentioned that the visuals \textit{"even though we were doing different roles, it demonstrates what actually happened in that scenario pretty well."} The visualisations, capturing both communication and movement, were further commended. P9 noted in the communication interaction diagram (Figure \ref{fig:figure1}--C) that the dot sizes \textit{"reflects how much we were speaking."} Coupled with the ward map (Figure \ref{fig:figure1}--B), P9 found it showed that while the primary nurse spoke the most, they moved the least, concluding it \textit{"reflects physically moving around and communication fairly."}

\subsubsection{Inadequate Analysis and Visualisation Design}
A prominent theme was the lack of consideration in the decision-making process of data selection, analysis, reporting, and visualisation, raising concerns about fair representation. Two sub-themes were identified as follows: 

\textbf{\textit{Visualisation not Representing the Full Context \& Nuances.}} The limitations of data visualisation in capturing the depth and context of communication were highlighted by students (n=5). P11 remarked on the challenge of illustrating individual contributions, arguing that, \textit{"It's quite difficult to show how each team member participated or contributed to this scenario."} P5 noted that, as a secondary nurse, visualisations might misrepresent their level of engagement: \textit{"you can see that we don't communicate a lot because we were doing that task."} P8 critiqued the communication behaviour diagram (Figure \ref{fig:figure1}--D) for its lack of context, stating it merely \textit{"recognises that people were asking questions and giving answers, not actually [providing] any context of what they were [saying]."} P7 expressed reservations about the ward map, pointing out the data is in aggregate form as it does not \textit{"show how we moved minute by minute."} Furthermore, P6 emphasised the system's limitations in capturing the \textit{"quality and effectiveness of communication"}, for example, \textit{"a nurse was talking but they were taking quite some time to get to their point... which would be contributing to the ward map and communication interaction."}, and the lack of recognition for \textit{"each person's communication style"}, whether they are more passive, active, or direct.

\textit{\textbf{Individual Contributions \& Role Differentiation}} Differentiating individual contributions during simulations, especially between primary and secondary nurses, proves challenging. P3 noted the communication behaviour diagram mainly shows team interactions, stating: \textit{"The communication behaviour… you can [only] see if the team overall did a lot of acknowledging and questioning. It might just be two people going back and forth. So that's harder to differentiate each role."} From a role perspective, P10 observed: \textit{"I guess the primary nurses are more expected to do task allocation and escalation, whereas it's... more expected for the job of acknowledging and questioning to be for the secondary nurses."} P10 further suggested it would be beneficial to see \textit{"which roles were doing most of those communication behaviours."} P6's comments highlighted potential biases, noting that one might assume \textit{"red [with bigger circle] is doing a lot… they deserve a higher grade compared to green and yellow."} while the ward map suggests \textit{"green and yellow is not moving around a lot. That means they're not doing anything."}

\subsubsection{Flawed or Incomplete Data}
Another theme from the interviews was the misrepresentation of information, mainly attributed to missing data or technology flaws. P7 observed inconsistencies between their perceived communication levels and the visual representation, stating, \textit{"I think we talked a lot... in my opinion, my arrows from me as a Primary Nurse 1 to Primary Nurse 2 should be bigger."} Similarly, P13 pointed out the visual omission of a secondary nurse's communication, mentioning that \textit{"having no arrows coming out shows that they weren't communicating to anyone other than the patient, which I do not think was accurate."} P5 suggested these discrepancies might be due to \textit{"technical difficulties, like if there was some problem with the mic."} Additionally, the exclusion of non-healthcare professionals like patient's relatives impacts the accurate interpretation of events. P4 noted that the system failed to capture their \textit{"interaction with the patient's relative in Bed 3."} With no data for the patient's relative, the ward map suggested inactivity, making it seem \textit{"like we weren't doing anything."} In an extreme case, P2 highlighted the unfairness of evaluating based on incomplete data, pointing out the complete omission of Secondary Nurse 1 from the ward map, \textit{"She doesn't even exist there, so I feel like it's not fair in that aspect."}

\subsubsection{MMLA Visualisations for Assessment}
In this sub-section, we report students' views on the use of MMLA visualisations for \textit{fair assessment} of individual and team performance. Students' opinions were divided. Only P3 supported the idea without any strings. Some students (n=5; P1, P4, P10, P12, P13) felt that MMLA visualisations could be used fairly for assessment under certain conditions, while the majority (n=8; P2, P5-9, P11, P14) believed it would be unfair. Beyond the aforementioned themes, students provided suggestions to enhance the visualisations for assessment. Many highlighted the need to incorporate video and audio data. P2 viewed them as a \textit{"collection of memory"}, and P4 felt they could \textit{"help better understand what's happening in the scenario."} P9 emphasised their ability to \textit{"capture body language."} Other suggestions included multiple trials and role rotations, as proposed by P7 and P8, to identify consistent performance trends and enhance assessment accuracy.

\subsection{Accountability (RQ2)}
In this subsection, we focus on data security as a lens to investigate students' perceived accountability in data access, potential misuse and consequences, as well as the measures to mitigate such risks.

\subsubsection{Data Access}

We summarise the results by the types of stakeholders mentioned, namely students (e.g., whole and other classes), educators (e.g., coordinators and teachers), and third parties (e.g., researchers and admins). 

\textit{\textbf{Students}} 
A majority of the participants (n=12) believed that the whole class should have access to their data, primarily for self-reflection and collaborative learning purposes. P7 saw the benefits in self-improvement, stating that students could \textit{"reflect on their own performance"} and identify areas of strength or needed growth. P6 felt the data should be available \textit{"right after the simulation just for reflection"}, helping students quickly understand their actions. P5 noted the data's value in reviewing team dynamics. However, P3 and P11 felt access should be limited to the four students participating in the same simulation scenario. For example, P3 mentioned, \textit{"not necessarily the whole class... you were able to reflect on your experience without comparing."} Only six participants felt it would be appropriate to share data access with students from other classes. The main concern, as voiced by P6 and P13, was fostering competitive or negative attitudes and the risk of judgment from peers. P10 expressed concerns about comfort, noting some students might be \textit{"scared of judgement from other people"}, which could discourage them from participating actively. On the other hand, some participants, like P4 and P14, saw data sharing as a learning opportunity. P4 believed younger students, in particular, could benefit as they might not have had much hands-on experience, suggesting it could be valuable for those who \textit{"haven't had that much experience on placement and other scenarios."} 

\textit{\textbf{Educators}}
Most participants (n=10; n=13) supported granting data access to the unit coordinator and other teachers to enhance teaching and assessment methods. P13 saw the visualisations as an \textit{"overall snapshot"} that could highlight areas for teaching improvement, which could serve as a \textit{"reflection tool for future courses"}. While P2 and P6 highlighted the value for teachers assessing student performance, P7 believed teachers could use the data to pinpoint student \textit{"weakness"}, but was hesitant about sharing with the unit coordinator due to concerns about fairness in assessment. Some participants, like P8 and P11, were cautious about unrestricted access. They preferred differentiating access levels between video, audio, and visualisations. P8 felt teaching staff should be among the \textit{"few people"} with access to video and audio, whereas others should see only anonymised visualisations. P11 advocated for a more limited approach, suggesting staff access only the graphs to avoid potential bias or judgement, for example, \textit{"one student was having a bad day, or they're really stressed just before the scenario, they may perform poorly, and that may pose some judgement."} 

\textbf{\textit{Third Parties}}
Four participants emphasised the value of researchers accessing their data. P6 outlined the broader potential of such data, suggesting it might be \textit{"useful in the general sense of a study like what you're doing right now. If it's proved to be efficient and accurate, it could also go further into a health organisation's research and study and overall in our country. And potentially, if it gets that far could be international as well."} Only three students felt that university administrative staff need their data access. P3 reasoned that the admin team should \textit{"have their hands on everything"} to support teaching staff effectively. P10 expressed trust in the institution's protocols and the integrity of administrative staff, believing they would uphold student privacy and adhere to the organisation's code of conduct. They felt assured that measures would be in place to protect students, stating, \textit{"I would trust staff would have procedures in place to keep students safe."}

\subsubsection{Misuse of Data and consequences}

Four sub-themes are summarised as follows:

\textit{\textbf{Judgement \& Critique}} A primary concern among participants was the potential for judgement and critique from others, including students, teachers, or outsiders without a nursing background. P5 emphasised the simulation as \textit{"a safe place"} for learning and worried that outsiders might misunderstand situations and use them negatively. P6 raised concerns about potential misrepresentation, where outsiders might create a \textit{"false story"} and link it to unrelated adverse medical outcomes, so that\textit{ "someone might make up a false accusation of a particular hospital or particular person."} This participant also recalled an instance where past student videos were used in teaching, leading to critiques, which made them wary about their own data being \textit{"potentially critiqued"}. P1 expressed concerns that students could be \textit{"disempowered or lose motivation to pursue learning"}, compromising their learning experience and outcomes. 

\textit{\textbf{Privacy Concerns}} Many participants voiced concerns about the exposure of personal identity, especially from video and audio collected during the simulation.  P4 articulated the discomfort as follows: \textit{"I would be concerned, 'cause I don't feel comfortable having my face... exposed to others who aren't meant to have access to it."} P7 pointed out the problem, in essence, summarising,\textit{ "because our pictures are there. So people can identify us easily."} Moreover, P4 and P5 pinpointed their perceived nature of data misuse as \textit{"breach of privacy"} and \textit{"breach of confidentiality"}. Lastly, P2 warned about the danger of using Artificial Intelligence (AI) for malicious intent. \textit{"I think nowadays like AI... is on a very big hike. So I think it could make, for example, an AI of me saying something. [Because] my face is there, my voice is there"}, and remarked, \textit{"Technology nowadays is scary."} 

\textit{\textbf{Impact on Future Career}} Some students also stressed the potential consequences on their employment prospects if data were made accessible to future employers. P8 conveyed the anxiety that \textit{"what if your future employer... [has] a look at your performance?"} Such concern can be linked to judgment. As P11 suggested, \textit{"it could be misused in regards to judging a student's or a participant's performance when they didn't know they would be judged in a way... A potential employer might use that and discriminate against us."} P10 further elaborated that organisations might find fault in students' actions from the video and audio, which may affect \textit{"their willingness to employ us later."} 

\textit{\textbf{Academic Integrity}} Finally, students are worried that leaked videos could compromise the authenticity of subsequent learning scenarios, leading to academic integrity issues. P5 pointed out that if students from other classes view the simulations in advance, \textit{"it's not an actual reaction anymore."} P4 validated this sentiment by imagining another scenario where \textit{"a past student has given a current student access to the videos and the visualisations so that they can get a better score... they've breached academic integrity and policies."}

\subsubsection{Delineation of Accountability}
Students delineated accountability for data misuse between institutional and individual levels. Institutions, particularly universities and research teams, are seen as the primary custodians of data, bearing the responsibility for its safeguarding and ethical handling. P2 stressed the role of the \textit{"Director of the program, the uni as a whole [is] responsible... they should really ensure that this sort of stuff is encrypted."} P12 reiterated this trust in institutions, believing the university \textit{"should be responsible."} P7 emphasised the importance of the research team and university implementing consistent privacy policies, suggesting that anyone handling the data should protect the privacy of those in the videos, especially when sharing with third parties. At the individual level, responsibility for data extends to all who interact with, access or disseminate the data. Teachers, as the bridge between data and its pedagogical implications, were particularly highlighted. P1 believed teachers \textit{"should be responsible for like how they communicate with the student"} and their conveyed judgements. Additionally, any individuals, including students, who misuse the data, either deliberately or accidentally, should be held accountable. P4 pinpointed that \textit{"the person who gave the information, as well as a student who had the intent to use it for to cheat essentially would get in trouble."} P9 further emphasised the seriousness of unauthorised data sharing, stating that \textit{"whoever would be sharing this video... should have to suffer the consequences for it."}

\subsubsection{Prevention of Data Misuse}
Participants provided insights into the measures that should be in place to protect data from misuse. First, participants like P2 highlighted strict access control, noting the importance of \textit{"encryption"} coupled with \textit{"limiting access"} to the data. P9 further elaborated that students should access the data \textit{"indirectly"} where they \textit{"saw once on the screen"}, with P7 suggesting to \textit{"see the videos on a centralised platform"} for a limited time. Moreover, P10 suggested measures like ensuring a secure online environment, relying on existing platforms like Moodle, while P11 introduced a novel concept of using \textit{"blockchain encryption"}, inspired by its application in healthcare, as a potential measure to ensure data integrity. Lastly, P3 emphasised anonymity, suggesting that videos should maintain "\textit{the quality isn't good enough to do facial recognition}" yet retain enough detail for those in the simulation to recognise their participation.

\subsection{Transparency (RQ3)}

\subsubsection{Data Collection \& Tools} Students' understanding of data capturing and the tools used was a prominent theme. Participants noted the multimodal data collection methods in the simulation. P2 summarised the multimodal nature of the tools, stating, \textit{"I do understand that the belly bag was for the sake of positioning, the microphone for the sake of audio obviously, and the cameras in the room also help with the positioning."} P12 linked the tools directly to the visualisations, observing how the microphones and belly bag collected data for the ward map. However, not all students clearly understood how and what data was collected. P4 assumed data was mainly from \textit{"audio and video"}, while P6 believed cameras played a primary role in \textit{"generating the ward map and prioritisation behaviour chart [Figure \ref{fig:figure1}--A]."} In fact, the positioning data for both visualisations were captured by the positioning sensor inside the belly bag. This misconception was evident when P7 questioned the communication interaction diagram, wondering \textit{"how you really know to whom we talk."} There was a gap between students' perceived and actual understanding. P9, discussing the prioritisation chart, felt the combination of tools captured data, even though the chart relied solely on the indoor positioning system. Despite this, the student believed their understanding was \textit{"adequate, maybe not a hundred percent."}

\subsubsection{Data Processing and Analysis} While most participants (n=11), except P1, P2 and P14, attempted to explain their understanding of data processing, many relied on guesswork. Several believed data was synthesised to create the visualisations. P3 presumed, \textit{"All of my data would be put into one thing, and then it would be shuffled around like in Excel, and then you create graphs."} P5 thought the system \textit{"gathers the thoughts from the simulation, and grabs it into the different visualisations"}, highlighting a general lack of clarity around data processing. The complexity and volume of data may have contributed to this confusion. P11 observed that while the collected information seemed \textit{"complicated"}, its presentation was \textit{"rather easy to interpret."} P12 was puzzled by the rapidity of data processing that \textit{"how exactly that was transformed so quickly"}, noting that the visualisations appeared almost immediately post-simulation. A lack of data and AI literacy could be another factor. Some students, like P6 and P13, believed data was manually analysed by researchers. P6 thought that the creation of a communication behaviour diagram required human interpretation, while P13 assumed researchers filled hexagons on the ward map based on student movement and speech. In contrast, P7 suggested that \textit{"maybe AI"} was involved, while P11 envisioned an automated process with \textit{"an algorithm processing positioning data and transcribing"} speech.

\subsubsection{Motivation to Learn} Expanding on these misconceptions and lack of understanding, we inquired students about their motivations for learning more details about data collection and processing. Out of the 14 students, only P3 showed no interest as long as they could utilise and reflect upon the visualisations. In contrast, P8 found \textit{"it's interesting...helps me understand the charts a bit better."} Further, P13 felt \textit{"seeing how it [visualisation] was achieved"} could help them to \textit{"better understand how accurate it is."} Notably, P14 highlighted that understanding \textit{"how the data was processed" can contribute to "how much you'll be able to trust those [visualisations].}"

\subsection{Ethics (RQ4)}
This subsection focuses on investigating and making sense of students' decisions and preferences regarding consenting methods. We present the findings in three parts, corresponding to the questions we asked in the interview:

\subsubsection{Informed Consent \& Explanatory Statement}

By inquiring about students' awareness and engagement with the explanatory statement, we examined their understanding of the provided information in the decision-making process. While all participants were aware of the statement and their rights to opt-out, only five (P5, P9, P10, P12, P14) fully read it. The remaining nine had various reasons for not doing so. P1 expressed a carefree attitude, saying \textit{"I didn't read it or anything. But I didn't really mind if stuff [data] was used."} P3 was enthusiastic, noting \textit{"I didn't read it in full. But that's because I was already ready to engage and happy for this data [to be used]"}, suggesting trust in the data usage. P6 felt it was too lengthy, remarking \textit{"Because I never read the terms and conditions, it's just too long"}, despite that the document was merely over two pages. Additionally, reading the document did not guarantee full comprehension, with P10 admitting they \textit{"skimmed over it pretty quickly"} and P14 not reading \textit{"it word by word."}

\subsubsection{Motivation of Participation}
We also explored the motivation for students to stay in the study and allow their data to be used for research. A prominent factor was the belief in the values and benefits of their contribution to research, education, and even practice. P1 trusted the authenticity of the research, noting \textit{"If you guys are doing it for research, I know that you'd be doing it for something cool or good."} P2 believed their participation might \textit{"benefit someone [a PhD or Master's student] and help them graduate."} P8 saw the potential for their data to benefit the developments of \textit{"better education and learning tools"}, and P12 initially felt \textit{"daunting at the start because I knew everybody's watching"}, but later believed the data could enhance future practices. P3 viewed their participation as both a personal learning opportunity and a means to improve patient care, hoping that \textit{"this [participation] is one little thing that I get benefit out of, and a lot of other people can get benefit out of."} Trust in the institution also played a role. P11 expressed \textit{"I did trust the university to use my data in a way that is responsible."} Similarly, P10 felt \textit{"comfortable" }that the data would not be used to judge them.

\subsubsection{Opt-out Versus Opt-in Consent Method}

When discussing the preference between opt-out and opt-in consenting, 11 participants (P1-4, P6, P8-10, and P12-14) favoured the opt-out approach used in the study. In contrast, three participants (P5, P7, and P11) preferred the opt-in method. For those preferring the opt-out method, the primary reason was the ease of participation and little effort required by this approach. P12 found it \textit{"easier"}, and P1 believed many students tend to avoid extra steps, observing \textit{"knowing my cohort, people are kind of lazy... if you want data, give them the choice to opt out."} P6 felt that \textit{"if I need to opt in, I probably wouldn't have been participating"}. This student also suggested that employing the opt-in method would require \textit{"more effort"} to recruit the same number of participants whereas using the opt-out approach, \textit{"if there are people that disagree with their data being collected, they would take action."} Similarly, P13 anticipated a higher participation rate that \textit{"with the opt-out approach, maybe get more people to be using the study and it's still made very clear to the people who definitely don't want to be in it, that it is an option. So I just reckon it works better." } Nonetheless, some pinpointed the importance of clear communication about their options to exit the study prior to the simulation. P10 advocated for an \textit{"informed opt-out"} to ensure students know that \textit{"they can opt out and what it means if you don't opt out.}" P4 reinforced the importance, expressing that confusion between participating in the simulation and the research was not clarified \textit{"until right before when I got given the headset and the belly bag."} Conversely, proponents of opt-in valued the autonomy and control it offered. P7 appreciated that it gave people \textit{"autonomy"}, and P11 sought \textit{"more control and more informed choice"}. P5 felt the opt-in method would prevent misunderstandings, ensuring students knew they were \textit{"participating in this study"}.

\section{Discussion}

In this study, we uncovered students' perceptions of FATE issues of MMLA visualisations in an authentic learning setting. For \textbf{fairness (RQ1)}, the majority of students equated fairness with the accuracy of visualisations, aligning with previous studies on LA dashboards which emphasised the equitable and impartial reporting and visualisation of LA \cite{verbert2020learning}. This accurate reporting could be particularly challenging for MMLA. Apart from ensuring algorithmic fairness, MMLA, reliant on sensing technologies, can encounter technical challenges leading to \textit{flawed or incomplete data} during the collection process, leading to biased and unfair representation, echoing with prior logistical concerns \cite{yan2022scalability,ouhaichi2023rethinking}. Likewise, students also attributed their concerns to the lack of consideration in the decision-making process during data selection, analysis, reporting, and visualisation, resonating with prior works on algorithmic fairness \cite{jiang2021towards, kim2022information} that understanding and addressing fairness issues is a holistic and multi-stage process \cite{barbierato2022methodology}. Specifically, in MMLA, selective data handling could lead to potential biases such as \textit{the lack of depth and context of communication} and \textit{the failure to account for individual contributions}. Additionally, students' perceived fairness of the visualisations is also context-dependent, emphasising the need to differentiate the individual challenges, for example, setting role-based rubrics when their grade is at stake. As the goal of LA is to optimise learning, fairness should not be limited to algorithms but also extend to a student-centric process where the decisions related to reporting and visualisation are also considered \cite{verbert2020learning}. 

Regarding \textbf{accountability (RQ2)}, most students supported sharing their data within their class for collaborative learning and reflection but were cautious about sharing with other classes, fearing judgment. Educators, particularly unit coordinators and teachers, were largely supported to access the data to improve teaching methods, though some students advocated for differentiated access levels. For third parties, some participants recognised the potential value for researchers, while only a few felt university admin staff should have access. These views align with \citeauthor{mangaroska2021challenges}'s \citeyear{mangaroska2021challenges} findings, showing students' willingness to share anonymous data with educators. However, concerns were raised about potential misuse, including judgment, privacy breaches, impact on future career prospects, and academic integrity. These issues are particularly evidenced in identifiable personal data such as audio and video, echoing past studies \cite{mangaroska2021challenges, prinsloo2023multimodal, liuunderstanding}. Lastly, students believed accountability for data misuse should lie both with institutions like universities and research teams, and individuals, especially teachers and those misusing the data. These findings resonate with prior works that accountability is beyond an individual person but distributed across multiple parties \cite{prinsloo2017elephant, pardo2014ethical}. 

For \textbf{transparency (RQ3)}, while participants understood the multimodal nature of tools, they held misconceptions about data capture specifics. Many students showed uncertainty about data processing and analysis, with varied guesses ranging from manual analysis by researchers to automated AI systems. Such misconceptions and lack of understanding resonate with prior studies of transparency in higher education \cite{tsai2020empowering}. Specifically, transparency might be more challenging to achieve in students than in teachers, where teachers may have more theoretical and contextual knowledge for comprehending the data analysis \cite{kasepalu2021teachers}. Yet, most students were keen to understand the data collection and processing as such information could foster their trust in the technology and visualisations. This finding suggested the need to develop measures and mechanisms to evaluate and cultivate students' comprehension of the data collection and analysis processes, especially in MMLA studies where multiple sensors that might be unfamiliar to students could be used to capture physical or physiological data \cite{cukurova2020promise, prinsloo2023multimodal}. Similar to the transparency index for AI in education \cite{chaudhry2022transparency}, the field of MMLA also needs a transparency evaluation framework to account for the unique challenges of this field.

In terms of \textbf{ethics (RQ4)}, specifically examining the consenting methods, all participants were aware of the explanatory statement, but only five fully read it due to reasons ranging from carefree attitudes to perceptions of lengthiness. These perceptions align with prior concerns regarding consenting without full comprehension \cite{prinsloo2016student, li2022disparities}. This finding supports the imperative need to scrutinise our conception of informed consent from a dichotomous view to a continuous scale that captures different levels of comprehension and develop practical tools to measure such differences like an informed consent comprehension test \cite{beardsley2020enhancing}. Most participants favoured the opt-out approach for its ease, anticipating a higher participation rate. However, the importance of clear communication and understanding of their options was emphasised. Conversely, those preferring the opt-in method valued the greater autonomy and control it provides, ensuring participants are more informed and conscious of their choice to participate. This preference for the opt-out approach resonates with prior works, which also reported similar results \cite{junghans2005recruiting, li2019impact}. Regardless of the consent method, clear communication and ensuring autonomy are essential for all participants, further supporting the need to transform the current consenting practices to ensure informed participation in LA and MMLA initiatives \cite{li2022disparities, beardsley2020enhancing, sun2019s}. 

The findings have several \textbf{implications} for future research and practices. Firstly, there is a need to ensure the fairness of visualisations by addressing algorithmic, technical, and representation challenges \cite{verbert2020learning, ouhaichi2023rethinking, barbierato2022methodology}. This necessitates a student-centred approach that prioritises transparent and equitable data representation and decision-making process, ensuring the alignment between researchers' selective representation of learner data and students' learning needs \cite{gavsevic2015let}. Secondly, while students appreciate the benefits of sharing data for reflective and collaborative learning, they remain cautious about the broader dissemination, particularly when personal identifiers are involved. This highlights the importance of differentiated access levels and robust protocols to safeguard students' data, where different parties can be held accountable in case of data misuse \cite{prinsloo2017elephant, pardo2014ethical}. Furthermore, the evident misconceptions about the data collection and analysis processes underline the urgency of fostering transparency \cite{tsai2020empowering}, potentially through developing transparency evaluation frameworks for MMLA and tailoring educational interventions to cultivate student comprehension \cite{chaudhry2022transparency}. Lastly, the complexities surrounding informed consent warrant a shift from traditional dichotomous views to more nuanced, continuous scales that accurately reflect comprehension levels. As such, future studies should prioritise the development of mechanisms that facilitate in-depth understanding, coupled with flexible consent options that respect student autonomy \cite{beardsley2020enhancing, li2022disparities}. Collectively, these insights motivate a call to action for the LA and MMLA community to devote more research initiatives to focus on FATE, ensuring that the field remains student-centred and ethically robust.

Several \textbf{limitations} should be considered when interpreting the current results. Firstly, FATE is a multi-dimensional concept; in fact, each aspect of FATE is a multifaceted concept that covers a range of implications from technical to socioeconomic. Consequently, our exploration might not capture all the nuances and complexities. We predominantly focused on students' perceptions, which, while insightful, may not represent the broader stakeholders' views, such as educators, administrators, and tech developers. Secondly, the context of the study, being situated in an authentic collaborative learning setting and as part of formative assessment, may not be generalisable to other educational contexts, such as during self-regulated learning or in summative assessments where students' grades were at stake. Thirdly, students' perspectives on FATE may also limited to the sensors and technology we deployed. For example, there could be differences in students' perspectives when more personal data, such as heart rate, were used for the visualisations, where they might be more hesitant to share such data \cite{mangaroska2021challenges}. Lastly, there is a selection bias, where we only investigated the perspectives of a sample of students who were willing to participate. Future studies should consider using a more diverse set of methodologies and including a broader range of participants to provide a more comprehensive understanding of FATE issues in LA and MMLA.

\section{Conclusion}

In this study, we delved into students' perceptions surrounding FATE issues of MMLA visualisations within an authentic collaborative learning setting. The findings highlight the significance of ensuring fairness in data visualisation, the nuanced complexities surrounding data sharing, the imperative of transparency in data processes, and the ethical intricacies of informed consent. While students value the benefits of MMLA, they also emphasise the importance of ethical considerations, highlighting a pressing need for the LA and MMLA community to actively investigate and address FATE issues. As technology and educational paradigms continue to advance, it becomes critical for LA researchers and practitioners to prioritise a student-centric approach, fostering a balance between innovation and ethical responsibility.

\phantomsection

\section*{Declaration of Conflicting Interest} 

\addcontentsline{toc}{section}{Declaration of Conflicting Interest} 

The author(s) declared no potential conflicts of interest with respect to the research, authorship, and/or publication of this article.

\section*{Funding} 

\addcontentsline{toc}{section}{Funding} 

This research was funded partially by the Australian Government through the Australian Research Council (project number DP210100060). Roberto Martinez-Maldonado’s research is partly funded by Jacobs Foundation.
\phantomsection
\bibliography{reference}


\end{document}